\documentclass{article}%
\usepackage[top=3cm, bottom=3cm, left=3cm, right=3cm]{geometry}
\usepackage{amsmath}
\usepackage{amsfonts}
\usepackage{amssymb}
\usepackage{graphicx}%
\begin{document}

\title{ Phenomenology of light mesons within a chiral approach}
\author{F. Giacosa$^{1}$, D. Parganlija$^{1,2}$, P. Kov\'{a}cs$^{1,3}$, and Gy.
Wolf$^{3}$\\$^{1}$\emph{Institute for Theoretical Physics, }\\\emph{Johann Wolfgang Goethe University,}\\\emph{Max-von-Laue-Str. 1, D-60438 Frankfurt am Main, Germany}\\$^{2}$\emph{Institute for Theoretical Physics, Vienna University of
Technology,}\\\emph{Wiedner Hauptstr. 8-10, A--1040 Vienna, Austria}\\$^{3}$\emph{Institute for Particle and Nuclear Physics, }\\\emph{Wigner Research Center for Physics, Hungarian Academy of Sciences, }\\\emph{H-1525 Budapest, Hungary }}
\maketitle

\begin{abstract}
The so-called extended linear sigma model is a chiral model with
(pseudo)scalar and (axial-)vector mesons. It is based on the requirements of
(global) chiral symmetry and dilatation invariance. The purpose of this model
is the description of the hadron phenomenology up to 1.7 GeV. We present the
latest theoretical results, which show a good agreement with the experiment.

\end{abstract}

In this paper\footnote{Based on the presentation of F.G. at the 12th
International Workshop on Meson Production, Properties and Interaction,
KRAK\'{O}W, POLAND, 31 May - 5 June 2012.} we describe a chiral $\sigma$
model, called `extended Linear $\sigma$ model (El$\sigma$m)', in which scalar,
pseudoscalar, vector, axial-vector quark-antiquark mesons and, in addition, a
scalar dilaton/glueball field are the basic degrees of freedom. The aim is to
develop a model with the symmetries of QCD which can describe the vacuum
phenomenology up to 1.7 GeV. The Lagrangian of the model is built by requiring
(i) global chiral symmetry and (ii) dilatation invariance. Although chiral
models are studied since long time \cite{geffen}, the here presented
El$\sigma$m represents the first attempt to treat in a unified chiral
framework (pseudo)scalar mesons (including the glueball) as well as
(axial-)vector ones. (Previous studies \cite{ko} exist only for $N_{f}=2$ and
not all the mentioned d.o.f. were taken into account.) It turns out that the
inclusion of (axial-)vector d.o.f. has a very strong influence on the overall
phenomenology, influencing also the decays in the (pseudo)scalar sector.

The explicit form of the Lagrangian in the mesonic sector reads (for a generic
number of flavors $N_{f})$ \cite{denis,stani,dick,dynrec}:%

\begin{align}
\mathcal{L}_{El\sigma m} &  =\frac{1}{2}(\partial_{\mu}G)^{2}-V_{dil}%
(G)+\mathrm{Tr}\left[  (D^{\mu}\Phi)^{\dagger}(D_{\mu}\Phi)-aG^{2}\Phi^{\dag
}\Phi-\lambda_{2}\left(  \Phi^{\dag}\Phi\right)  ^{2}\right]  -\lambda
_{1}(\mathrm{Tr}[\Phi^{\dag}\Phi])^{2}+\nonumber\\
&  +c_{1}(\det\Phi^{\dag}-\det\Phi)^{2}\text{ }+\mathrm{Tr}[H(\Phi^{\dag}%
+\Phi)]-\frac{1}{4}\mathrm{Tr}[L^{\mu\nu}{}^{2}+R^{\mu\nu}{}^{2}]\nonumber\\
&  +\frac{b}{2}G^{2}\mathrm{Tr}[L^{\mu}{}^{2}+R^{\mu}{}^{2}]+\frac{1}%
{2}\mathrm{Tr}[\hat{\delta}(L^{\mu}{}^{2}+R^{\mu})^{2}]-2ig_{2}\left(
\mathrm{Tr}[L_{\mu\nu}[L^{\mu},L^{\nu}]]+\mathrm{Tr}[R_{\mu\nu}[R^{\mu}%
,R^{\nu}]]\right)  \nonumber\\
&  +\frac{h_{1}}{2}\mathrm{Tr}\left[  (\Phi^{\dagger}\Phi)(L_{\mu}^{2}+R_{\mu
}^{2})\right]  +h_{2}\mathrm{Tr}\left[  \Phi^{\dag}L_{\mu}L^{\mu}\Phi+\Phi
R_{\mu}R^{\mu}\Phi\right]  +2h_{3}\mathrm{Tr}\left[  \Phi R_{\mu}\Phi^{\dag
}L^{\mu}\right]  \text{ }+...\text{ ,}\label{lag}%
\end{align}
where $D^{\mu}\Phi=\partial^{\mu}\Phi-ig_{1}(L^{\mu}\Phi-\Phi R^{\mu})$ and
dots represent further terms which are unimportant in the evaluation of decays
and (on-shell) scattering lengths. Following comments are in order:

(i) The (pseudo)scalar quark-antiquark mesons are described by the matrix
$\Phi=\left(  S^{a}+iP^{a}\right)  t^{a}$ ($t^{a}$ are the generators of the
group $U(N_{f})$). The pseudoscalar states are the pion, kaon and the $\eta$
and $\eta^{\prime}$ mesons. The assignment of scalar states is controversial
\cite{dynrec,amslerrev,varieglue} and represents one of the motivations of our
study. It turns out that the best agreement with the experiment is obtained
when the quark-antiquark scalar states of the model are assigned to the scalar
resonances between 1-2 GeV (theory in Refs. \cite{denis,stani,dick} and
experimental results in Ref. \cite{pdg}).

(ii) The (axial-)vector mesons are described by the matrices $L^{\mu
}=(V^{a,\mu}+A^{a,\mu})\,t^{a}$ and $R^{\mu}=(V^{a,\mu}-A^{a,\mu})\,t^{a}$.
The vector mesons are the usual $\rho,$ $\omega,$ $K^{\ast}(892),$ and $\phi$
mesons. The axial-vector mesons are assigned to $a_{1}(1230),$ $K_{1}(1270),$
$f_{1}(1285)$ and $f_{1}(1510)$.

(iii) The dilaton field is denoted by $G$ and its potential reads $V_{dil}(G)$
$=\frac{1}{4}\frac{m_{G}^{2}}{\Lambda_{G}^{2}}\left[  G^{4}\ln\left(  \frac
{G}{\Lambda_{G}}\right)  -\frac{G^{4}}{4}\right]  $ \cite{schechter}. The
parameter $\Lambda_{G}\sim N_{c}\Lambda_{QCD}$ sets the energy scale of the theory.

(iv) The $U(1)_{A}$ anomaly term is parametrized by the parameter $c_{1}$,
which has dimension [Energy]$^{4-2N_{f}}$.

(v) The matrix $H\propto diag\{m_{u},m_{d},m_{s},...\}$ describes explicit
symmetry breaking of both chiral and dilatation symmetries due to the bare
quark masses $m_{i}$. Similarly, in the (axial-)vector sector the diagonal
matrix $\hat{\delta}$ has been introduced.

(vi) Chiral symmetry breaking takes place when the parameter $a$ is negative.
In fact, upon the condensation of the field $G=G_{0},$ the `wrong' sign for
mesonic masses $aG_{0}^{2}<0$ sign is realized.

Once the shifts of the scalar fields $G\rightarrow G_{0}+G$ and $\Phi
\rightarrow diag\{\sqrt{2}\sigma_{N},\sqrt{2}\sigma_{N},...\}+\Phi$, where the
first term is a diagonal matrix with the quark-antiquark condensates, and
necessary redefinitions of the pseudoscalar and axial-vector fields have been
performed \cite{denis,dick}, the explicit calculations of physical processes
are lengthy but straightforward. (Note, the calculations are performed at
tree-level; the inclusion of loops is a task for the future, but only slight
changes are expected \cite{lupo}.)

\bigskip

In the following we summarize the results that we have obtained with the model
in Eq. (\ref{lag}):

$\bullet$ $N_{f}=2$ with frozen glueball $(m_{G}\rightarrow\infty)$
\cite{denis,denisproc}: by considering the limit $m_{G}\rightarrow\infty$ the
dilaton/glueball field is not an active d.o.f. One can neglect the gluonic
part of the Lagrangian of Eq. (\ref{lag}) and set $G=G_{0}=\Lambda_{G}.$ In
the works in Refs. \cite{denis,denisproc} it has been shown that the inclusion
of (axial-)vector mesons has a strong influence on the overall phenomenology.
For instance, the width of the scalar meson $\sigma$ (the chiral partner of
the pion) decreases substantially w.r.t. the case without (axial-)vector
mesons: for this reason, the identification of this field with the resonance
$f_{0}(500)$ is not favoured, because the theoretically evaluated width is
smaller than $200$\ MeV; this is at odd with the experiment, according to
which it is larger than $400$ MeV. On the other hand, the identification of
the $\sigma$ field with the resonance $f_{0}(1370)$ turns out to be in
agreement with the experimental results. The description of the (axial-)vector
resonances is also in agreement with the experiments reported in Ref.
\cite{pdg}.

$\bullet$ $N_{f}=2$ with active glueball $(m_{G}\sim1.5$ GeV$)$ \cite{stani}:
the glueball with a bare mass $m_{G}\sim1.5$ GeV, in agreement with the
lattice results \cite{latticeglueball}, has been investigated for the first
time in a chiral model with (axial-)vector mesons. (For other approaches see
Ref. \cite{varieglue} and refs. therein). The state $f_{0}(1500)$ results as
the predominantly ($75\%$) glueball state, and the rest of the phenomenology
is only slightly affected w.r.t. the previous case, in which $m_{G}%
\rightarrow\infty$. Moreover, also the gluon condensate has been evaluated and
is in agreement with lattice results.

$\bullet$ $N_{f}=3$ with frozen glueball $(m_{G}\rightarrow\infty)$
\cite{dick,nf3}: the results for this scenario are interesting because, for
the first time, it is possible to study in a chiral framework the overall
vacuum's phenomenology up to 1.7 GeV. The results for the best-fit scenario
are listed in Table 1. Eleven free parameters enter in the fit and the total
$\chi^{2}\simeq1$ signalizes a good agreement of the theory with the
experiment, see Ref. \cite{dick} for details. It is visible that the
resonances $a_{0}(1450)$ and $K_{0}^{\ast}(1430)$ are well described as
quark-antiquark fields. The scalar-isoscalar mesons are not included in the
fit, but can be studied as a consequence of it: the decay pattern and the
masses suggest that $f_{0}(1370)$ and $f_{0}(1710)$ are (predominantly) the
non-strange and strange scalar-isoscalar fields.

\begin{table}[th]
\centering
\begin{tabular}
[c]{|c|c|c|}\hline
Observable & Fit [MeV] & Experiment [MeV]\\\hline
$f_{\pi}$ & $96.3 \pm0.7 $ & $92.2 \pm4.6$\\\hline
$f_{K}$ & $106.9 \pm0.6$ & $110.4 \pm5.5$\\\hline
$m_{\pi}$ & $141.0 \pm5.8$ & $137.3 \pm6.9$\\\hline
$m_{K}$ & $485.6 \pm3.0$ & $495.6 \pm24.8$\\\hline
$m_{\eta}$ & $509.4 \pm3.0$ & $547.9 \pm27.4$\\\hline
$m_{\eta^{\prime}}$ & $962.5 \pm5.6$ & $957.8 \pm47.9$\\\hline
$m_{\rho}$ & $783.1 \pm7.0$ & $775.5 \pm38.8$\\\hline
$m_{K^{\star}}$ & $885.1 \pm6.3$ & $893.8 \pm44.7$\\\hline
$m_{\phi}$ & $975.1 \pm6.4$ & $1019.5 \pm51.0$\\\hline
$m_{a_{1}}$ & $1186 \pm6$ & $1230 \pm62$\\\hline
$m_{f_{1}(1420)}$ & $1372.5 \pm5.3$ & $1426.4 \pm71.3$\\\hline
$m_{a_{0}}$ & $1363 \pm1$ & $1474 \pm74$\\\hline
$m_{K_{0}^{\star}}$ & $1450 \pm1$ & $1425 \pm71$\\\hline
$\Gamma_{\rho\rightarrow\pi\pi}$ & $160.9 \pm4.4$ & $149.1 \pm7.4$\\\hline
$\Gamma_{K^{\star}\rightarrow K\pi}$ & $44.6 \pm1.9$ & $46.2 \pm2.3$\\\hline
$\Gamma_{\phi\rightarrow\bar{K}K}$ & $3.34 \pm0.14$ & $3.54 \pm0.18$\\\hline
$\Gamma_{a_{1}\rightarrow\rho\pi}$ & $549 \pm43$ & $425 \pm175$\\\hline
$\Gamma_{a_{1}\rightarrow\pi\gamma}$ & $0.66 \pm0.01$ & $0.64 \pm0.25$\\\hline
$\Gamma_{f_{1}(1420)\rightarrow K^{\star}K}$ & $44.6 \pm39.9$ & $43.9 \pm
2.2$\\\hline
$\Gamma_{a_{0}}$ & $266 \pm12$ & $265 \pm13$\\\hline
$\Gamma_{K_{0}^{\star}\rightarrow K\pi}$ & $285 \pm12$ & $270 \pm80$\\\hline
\end{tabular}
\caption{Best-fit results for masses and decay widths compared with experiment
(from Ref. \cite{dick}).}%
\label{Comparison1}%
\end{table}

$\bullet$ Other works related to the model: in Ref. \cite{gallas} the baryonic
part of the model has been presented for $N_{f}=2$ and in\ Ref. \cite{susagiu}
the nonzero density and nuclear matter saturation have been described. In Ref.
\cite{achim} part of the model has been investigated at nonzero temperature.

\bigskip

In the future we plan to perform the following investigations with the chiral
model in Eq. (\ref{lag}): (i) The case $N_{f}=3$ with $m_{G}\sim1.5$ GeV: the
inclusion of an active glueball represents the most straightforward extension
of our approach; the aim is a full description of scalar-isoscalar states
between 1-2 GeV, including also the resonance $f_{0}(1500)$ \cite{varieglue}.
(ii) The case $N_{f}=4$ (that is, including charmonia states) can be studied.
In this way we can test the implications of chiral symmetry also beyond the
low-energy sector. Note, only two additional parameters (both connected to the
charm mass) w.r.t. the case $N_{f}=3$ are needed. The Lagrangian is still the
one in Eq. (\ref{lag}). (iii) $\tau$ decays involving (axial-)vector mesons:
after including the weak-gauge bosons $W^{\pm},Z^{0}$ in the model the
spectral functions of vector and axial-vector states can be investigated. (iv)
Scalar states below 1 GeV: these states are not part of the model, but could
be added as tetraquarks along the line of Ref. \cite{tqmio} (see also Ref.
\cite{varietq} and refs. therein). (v) Inclusion of the pseudoscalar glueball
and the evaluation of its decays: this project can be relevant for future
experiments, such as as the PANDA experiment at FAIR. (vi) In-medium
properties and the phase diagram: the study of chiral phase transition
represents an important outlook of this work (for preliminary works see Refs.
\cite{susagiu,achim}). In fact, a great advantage of the linear realization of
chiral symmetry is the straightforward extension to nonzero temperature and
density. Moreover, also inhomogeneous condensates can be investigated.

\bigskip

\textbf{Acknowledgments:} Gy. Wolf and P. Kovacs thank Goethe University for
hospitality. They were partially supported by the Hungarian OTKA funds T71989
and T101438. The work of D. Parganlija and F. Giacosa was supported by the
Foundation of the Polytechnical Society Frankfurt. This work was financially
supported by the Helmholtz International Center for FAIR within the framework
of the LOEWE program (Landesoffensive zur Entwicklung
Wisschenschaftlich-\"{O}konomischer Exzellenz) launched by the State of Hesse.


\begin{thebibliography}{99}                                                                                               %


\bibitem {geffen}S.~Gasiorowicz and D.~A.~Geffen,
Rev.\ Mod.\ Phys.\ \textbf{41}, 531 (1969).


\bibitem {ko}P.~Ko and S.~Rudaz,
Phys.\ Rev.\ D \textbf{50}, 6877 (1994). M.~Urban, M.~Buballa, J.~Wambach,
Nucl.\ Phys.\ \textbf{A697 } (2002) 338-371. [hep-ph/0102260].

\bibitem {denis}D.~Parganlija, F.~Giacosa, D.~H.~Rischke,
Phys.\ Rev.\ \textbf{D82 } (2010) 054024. [arXiv:1003.4934 [hep-ph]].


\bibitem {stani}S.~Janowski, D.~Parganlija, F.~Giacosa, D.~H.~Rischke,
Phys.\ Rev.\ \textbf{D84 } (2011) 054007. [arXiv:1103.3238 [hep-ph]].


\bibitem {dick}
D.~Parganlija, P.~Kovacs, G.~Wolf, F.~Giacosa and D.~H.~Rischke,
arXiv:1208.0585 [hep-ph].


\bibitem {dynrec}F.~Giacosa,
Phys.\ Rev.\ \textbf{D80 } (2009) 074028. [arXiv:0903.4481 [hep-ph]].
F.~Giacosa,
AIP Conf.\ Proc.\ \textbf{1322 } (2010) 223-231. [arXiv:1010.1021 [hep-ph]].


\bibitem {amslerrev}C.~Amsler and N.~A.~Tornqvist,
Phys.\ Rept.\ \textbf{389}, 61 (2004).
E.~Klempt and A.~Zaitsev,
Phys.\ Rept.\ \textbf{454} (2007) 1 [arXiv:0708.4016 [hep-ph]].


\bibitem {varieglue}
C.~Amsler and F.~E.~Close,
Phys.\ Rev.\ D \textbf{53} (1996) 295 [arXiv:hep-ph/9507326].
W.~J.~Lee and D.~Weingarten,
Phys.\ Rev.\ D \textbf{61}, 014015 (2000). [arXiv:hep-lat/9910008];
F.~E.~Close and A.~Kirk,
Eur.\ Phys.\ J.\ C \textbf{21}, 531 (2001). [arXiv:hep-ph/0103173].
F.~Giacosa, T.~Gutsche, V.~E.~Lyubovitskij and A.~Faessler,
Phys.\ Rev.\ D \textbf{72}, 094006 (2005). [arXiv:hep-ph/0509247].
F.~Giacosa, T.~Gutsche, V.~E.~Lyubovitskij and A.~Faessler,
Phys.\ Lett.\ B \textbf{622}, 277 (2005) [arXiv:hep-ph/0504033].
F.~Giacosa, T.~Gutsche and A.~Faessler,
Phys. Rev. C \textbf{71}, 025202 (2005) [arXiv:hep-ph/0408085].
H.~Y.~Cheng, C.~K.~Chua and K.~F.~Liu,
Phys.\ Rev.\ D \textbf{74} (2006) 094005 [arXiv:hep-ph/0607206].
L.~Bonanno and A.~Drago,
Phys.\ Rev.\ C \textbf{79}, 045801 (2009) [arXiv:0805.4188 [nucl-th]].
V.~Mathieu, N.~Kochelev and V.~Vento,
Int.\ J.\ Mod.\ Phys.\ E \textbf{18} (2009) 1 [arXiv:0810.4453 [hep-ph]].


\bibitem {pdg}K. Nakamura \textit{et al}. (Particle Data Group), J. Phys. G
\textbf{37}, 075021 (2010).


\bibitem {schechter}A.~Salomone, J.~Schechter and T.~Tudron,
Phys.\ Rev.\ D \textbf{23}, 1143 (1981);
H.~Gomm and J.~Schechter,
Phys.\ Lett.\ B \textbf{158}, 449 (1985);
A.~A.~Migdal and M.~A.~Shifman,
Phys.\ Lett.\ B \textbf{114}, 445 (1982);


\bibitem {lupo}F.~Giacosa, G.~Pagliara,
Phys.\ Rev.\ \textbf{C76 } (2007) 065204. [arXiv:0707.3594 [hep-ph]].
F.~Giacosa,
arXiv:1110.5923 [nucl-th].


\bibitem {denisproc}
D.~Parganlija, F.~Giacosa and D.~H.~Rischke,
Acta Phys.\ Polon.\ Supp.\ \textbf{3} (2010) 963 [arXiv:1004.4817 [hep-ph]].
D.~Parganlija, F.~Giacosa and D.~H.~Rischke,
arXiv:0911.3996 [nucl-th].
D.~Parganlija, F.~Giacosa and D.~H.~Rischke,
PoS CONFINEMENT \textbf{8} (2008) 070 [arXiv:0812.2183 [hep-ph]].
D.~Parganlija, F.~Giacosa and D.~H.~Rischke,
AIP Conf.\ Proc.\ \textbf{1030} (2008) 160 [arXiv:0804.3949 [hep-ph]].




\bibitem {latticeglueball}Y.~Chen \textit{et al.},
Phys.\ Rev.\ D \textbf{73} (2006) 014516. [arXiv:hep-lat/0510074].


\bibitem {nf3}D.~Parganlija, F.~Giacosa, P.~Kovacs, G.~Wolf,
AIP Conf.\ Proc.\ \textbf{1343 } (2011) 328-330. [arXiv:1011.6104 [hep-ph]].
D.~Parganlija, F.~Giacosa, D.~H.~Rischke, P.~Kovacs, G.~Wolf,
Int.\ J.\ Mod.\ Phys.\ \textbf{A26 } (2011) 607-609. [arXiv:1009.2250
[hep-ph]].
D.~Parganlija,
Acta Phys.\ Polon.\ Supp.\ \textbf{4 } (2011) 727-732. [arXiv:1105.3647
[hep-ph]].


\bibitem {gallas}S.~Gallas, F.~Giacosa, D.~H.~Rischke,
Phys.\ Rev.\ \textbf{D82 } (2010) 014004. [arXiv:0907.5084 [hep-ph]].
S.~Gallas, F.~Giacosa, D.~H.~Rischke,
PoS \textbf{CONFINEMENT8 } (2008) 089. [arXiv:0901.4043 [hep-ph]].
S.~Wilms, F.~Giacosa, D.~H.~Rischke,
Int.\ J.\ Mod.\ Phys.\ \textbf{E16 } (2007) 2388-2393. [nucl-th/0702076].



\bibitem {susagiu}S.~Gallas, F.~Giacosa, G.~Pagliara,
Nucl.\ Phys.\ \textbf{A872 } (2011) 13-24. [arXiv:1105.5003 [hep-ph]].


\bibitem {achim}A.~Heinz, S.~Struber, F.~Giacosa and D.~H.~Rischke,
Phys.\ Rev.\ D \textbf{79} (2009) 037502 [arXiv:0805.1134 [hep-ph]].
A.~Heinz, F.~Giacosa, D.~H.~Rischke,
[arXiv:1110.1528 [hep-ph]].


\bibitem {tqmio}F.~Giacosa,
Phys.\ Rev.\ D \textbf{75} (2007) 054007 [arXiv:hep-ph/0611388].


\bibitem {varietq}
R.~L.~Jaffe,
Phys.\ Rev.\ D \textbf{15}, 267 (1977).
R.~L.~Jaffe,
Phys.\ Rev.\ D \textbf{15}, 281 (1977).
L.~Maiani, F.~Piccinini, A.~D.~Polosa and V.~Riquer,
Phys.\ Rev.\ Lett.\ \textbf{93} (2004) 212002 [arXiv:hep-ph/0407017].
F.~Giacosa,
Phys.\ Rev.\ D \textbf{74} (2006) 014028 [arXiv:hep-ph/0605191].
A.~H.~Fariborz, R.~Jora and J.~Schechter,
Phys.\ Rev.\ D \textbf{72} (2005) 034001 [arXiv:hep-ph/0506170].
A.~H.~Fariborz,
Int.\ J.\ Mod.\ Phys.\ A \textbf{19} (2004) 2095. [arXiv:hep-ph/0302133].
M.~Napsuciale and S.~Rodriguez,
Phys.\ Rev.\ D \textbf{70} (2004) 094043.
F.~Giacosa, G.~Pagliara,
Nucl.\ Phys.\ \textbf{A833 } (2010) 138-155. [arXiv:0905.3706 [hep-ph]].

\end{thebibliography}
\end{document}